\documentclass[aps,twocolumn,aip,superscriptaddress,amsmath,amsmath,amssymb,fleqn]{revtex4-1}
\usepackage{graphicx}
\usepackage{xcolor}
\usepackage[utf8]{inputenc} 
\usepackage[T1]{fontenc}    
\usepackage{hyperref}       
\usepackage{url}            
\usepackage{booktabs}       
\usepackage{amsfonts}       
\usepackage{nicefrac}       
\usepackage{microtype}      

\begin{document}
	
\title{Extinction-induced community reorganization in bipartite networks}

\author{Somaye Sheykhali}
\email{Author for correspondence: somaye@ifisc.uib-csic.es}
\affiliation{Instituto de F\'isica Interdisciplinar y Sistemas Complejos IFISC (CSIC - UIB), E-07122 Palma de Mallorca, Spain}


\author{Juan Fern\'andez-Gracia}
\affiliation{Instituto de F\'isica Interdisciplinar y Sistemas Complejos IFISC (CSIC - UIB), E-07122 Palma de Mallorca, Spain}

\author{Anna Traveset}
\affiliation{Instituto Mediterráneo de Estudios Avanzados IMEDEA (CSIC-UIB), E07121 Esporles, Spain }

\author{ V\'ictor M. Egu\'iluz}
\affiliation{Instituto de F\'isica Interdisciplinar y Sistemas Complejos IFISC (CSIC - UIB), E-07122 Palma de Mallorca, Spain}

\begin{abstract}
	We study how the community structure of bipartite mutualistic networks changes in a dynamic context. First, we consider a real mutualistic network and introduce extinction events according to several scenarios. We model extinctions as node or interaction removals. For node removal, we consider random, directed and sequential extinctions; for interaction removal, we consider random extinctions. The bipartite network reorganizes showing an increase of the effective modularity and a fast decrease of the persistence of the species in the original communities with increasing number of extinction events. Second, we compare extinctions in a real mutualistic network with the growth of a bipartite network model. The modularity reaches a stationary value and nodes remain in the same community after joining the network. Our results show that perturbations and disruptive events affect the connectivity pattern of mutualistic networks at the mesoscale level. The increase of the effective modularity observed in some scenarios could provide some protection to the remaining ecosystem.
\end{abstract}

\maketitle

\section{Introduction}

Mutualistic interactions between species are often represented as bipartite networks, where the interactions occur between two groups of species (generically resources and consumers), but not within the groups. Empirical mutualistic networks exhibit a number of macro-scale structural features such as nestedness\cite{Jordano,Gracia}, where specialists interact with proper subsets of the species that generalists interact with; modular organization, that captures the block structure \cite{Bascompte2010,Vazquez2005}; and stability, which can be measured as the largest eigenvalue of the appropriate matrix \cite{May1972}. Biological systems, and in general any complex system, are expected to withstand the loss of elements, either by random failure or driven by a directed perturbation (e.g., environmental change or a targeted attack) \cite{Burgos2007,Staniczenko2010}. In the context of ecology, loss of biodiversity as a consequence of environmental perturbations disrupts ecosystems and their functioning significantly.
The emergence of modularity is crucial for community ecology because such a compartmentalized structure can greatly influence dynamics, as the compartments buffer the spread of perturbation across the network \cite{Gardner,Gilarranz}. 

Here, we analyze how species extinction affects the structure of mutualistic networks. Our study focuses on consumer removals, as, as pollinators have a higher immediate extinction risk than plants and also loss of a pollinator species may cause the co-extinction of plants that depend on them \cite{Memmott,POTTS,Pedro}. We present a detailed analysis of the community structure in response to the loss of pollinator species, using an empirical mutualistic network.

The study of the changes in community structure under different extinction scenario sheds light on the fragility of ecological communities to species extinctions. We further emphasize our results by showing how in a model of mutualistic network growth \cite{Valverde} modularity and nestedness remain basically unchanged, in contrast to our results when extinction mechanisms are at play in a real bipartite network.

\section{Species extinction in empirical bipartite Networks}
We analyzed a plant-pollinator interaction network, sampled in Mallorca (Balearic Islands). The dataset was collected from a dune marshland located at sea level in the northeast of the island (Son Bosc; SB hereafter). The authors of \cite{Traveset} sampled insect-flower visitation events during the consecutive flowering season, from April to July on randomly selected flowering plants. A total of 696 flower visits between 80 plants and 162 pollinators were recorded (Fig. 1).

\begin{figure}[h!]
	\includegraphics[width=0.45\textwidth]{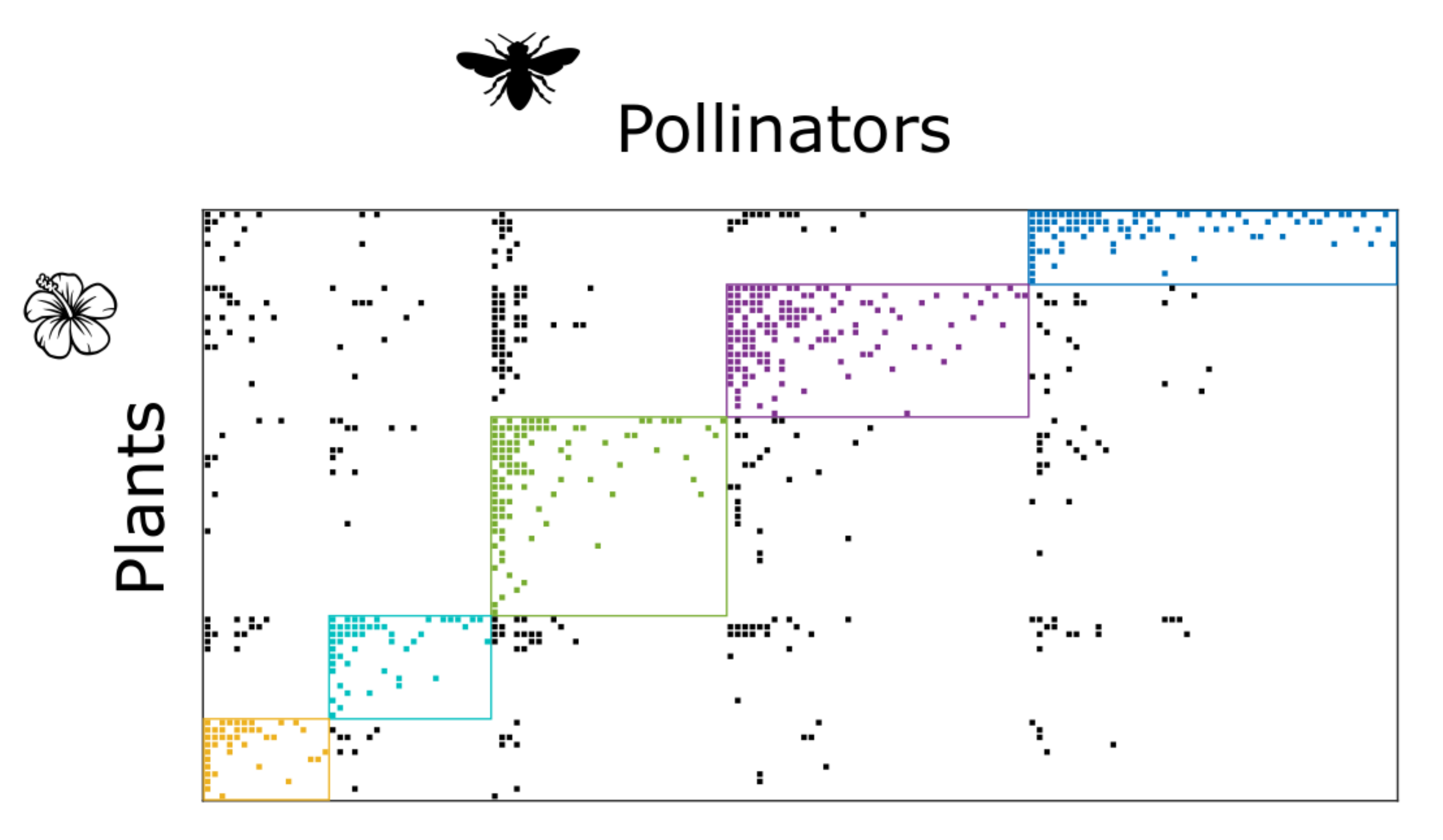}
	\caption{Bipartite incidence matrix depicting the community structure of a real plant-pollinator mutualistic network sampled in Mallorca (Balearic Islands). Plants are represented in the rows and the pollinator species in the columns. Mutualistic interactions inside a community are colored with the same color; black color was used for interaction across communities.}
	\label{WEB}
\end{figure}

The interactions found in general in mutualistic ecological communities are naturally represented with bipartite networks. These are composed of two different kinds of nodes:  resources, here the $n$ plants, and consumers, the $m$ insect pollinators. 
The interactions between resources and consumers are represented by the incidence matrix $A(n \times m)$, whose entries $A_{ij}$ are equal to 1 if there is a mutualistic relation between nodes $i$ and $j$, and $A_{ij}=0$ otherwise. In this work, we only consider the existence of an interaction but not its weight.

At each time step, an extinction event is modeled by removing a pollinator for node removal scenarios or an interaction for the interaction removal scenario. If a node loses all of its links, it becomes extinct. We simulate the loss of nodes and links with four different scenarios: (1) uniformly at random \cite{Albert}; (2) directed extinction, in which the removed pollinator is chosen with a probability proportional to her number of links (degree) \cite{Memmott,Evans,Gao}; (3) generalist scenario, in which pollinators are sequentially removed from the most to the least connected pollinator (in case of a draw one of them is chosen at random); (4) specialist scenario, in which nodes were sequentially removed from the least-degree pollinator to the most-degree \cite{Aizen}; and (5) random interaction extinctions, in which links were removed randomly to model the disappearance of an interaction. To balance for the different number of nodes and interactions, we measure time as the fraction of nodes removed or the fraction of links removed. That is, for scenarios where nodes are removed, each event represents a time step of $1/m$, while for interaction removal, a time step corresponds to $1/l$, where $l$ is the number of interactions. When all the nodes or all the links are removed, the time is equal to 1.
\begin{figure}[h!]
	\hspace{-0.4 cm}
	\includegraphics[width=0.5\textwidth]{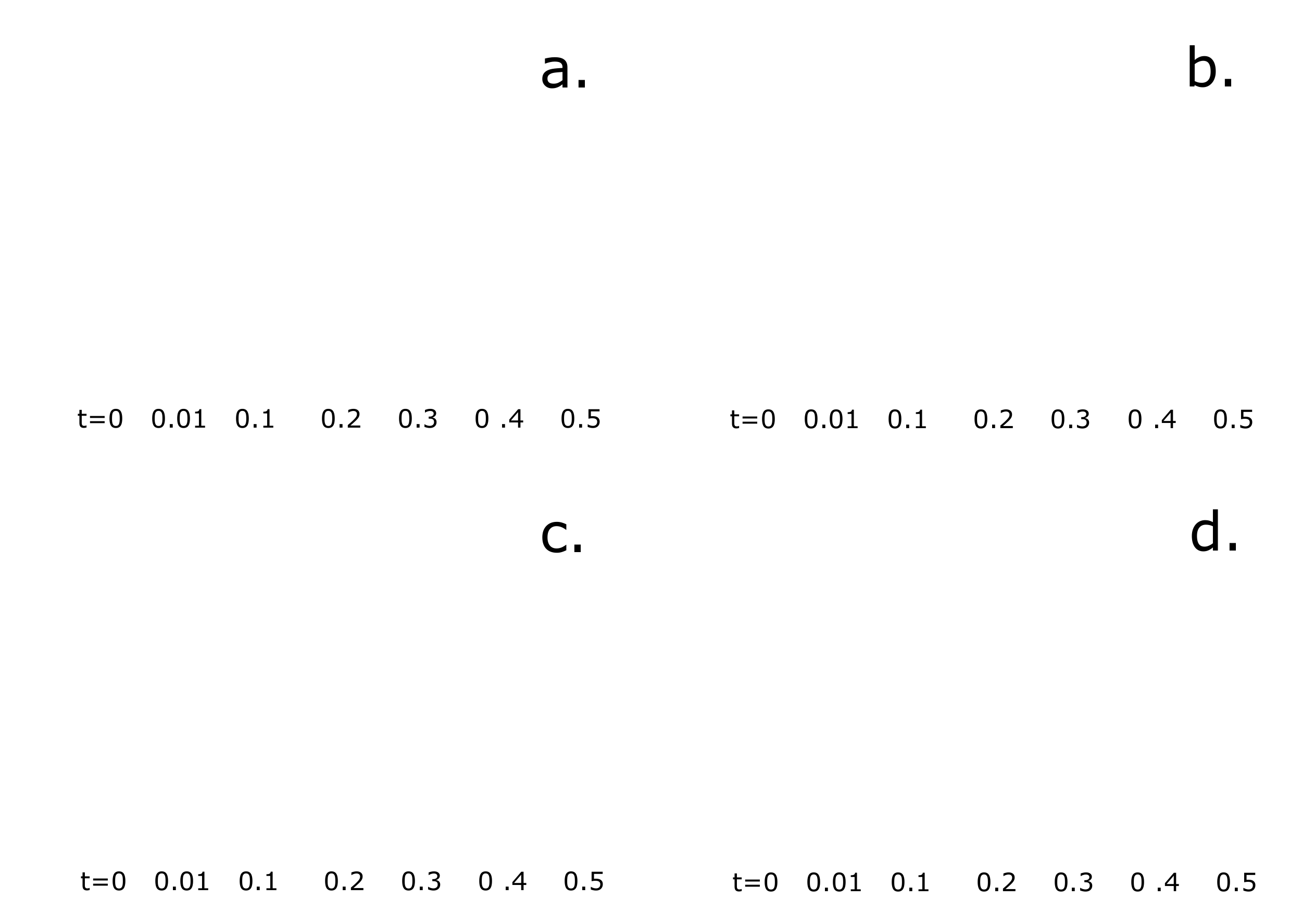}
	\caption{Time evolution of community structure after extinctions. Alluvial plots of SB data set under $(a)$. Random pollinator species removal. $(b)$ Generalist pollinator species removal $(c)$ Specialist pollinator species extinction. $(d)$ Random pollination interaction removal. Boxes show the communities at times 0.01, 0.1, 0.2, 0.3, 0.4 and 0.5.}
	\label{Alluvial}
\end{figure}

In order to compare visually the different extinction scenarios, we plot in Fig.~\ref{Alluvial} alluvial maps showing the community structure for a single realization of the different extinction dynamics. We have used the open-source library BiMat \cite{bimat} to compute modularity and community structure in bipartite networks. We show the resulting community structure for the same fraction of time as described above for all the scenarios. For random extinction and especially for specialist extinctions, the communities remain similar during the initial steps. This is in contrast with directed extinction and random interaction extinction, where the communities change much more from the initial condition. The dynamics of the community structure in the scenario  of directed extinctions behaves similarly as in the generalist extinction scenario (not shown). As expected, networks also lost more nodes in every level of directed extinction.

Note also that for a low fraction of extinction events directed extinction and interactions extinction perform very similarly. This is related to the fact that choosing an edge at random is similar to selecting nodes proportional to their degree.

We quantify these observations by measuring the modularity ($Q$) of detected communities, that is, densely connected non-overlapping subsets of nodes. The modularity of a bipartite network given a partition is defined as \cite{Barber}:
	\begin{eqnarray} 
	Q &&= \frac{1}{|{E}|} \sum_{i=1}^{n} \sum_{j=1}^{m} \left( A_{ij} - p_{ij} \right) \delta(g_i,h_j) \nonumber\\ 
	&&=\frac{1}{|{E}|} \sum_{i=1}^{n} \sum_{j=1}^{m} \left(  A_{ij} - \frac{k_i d_j}{|{E}|} \right) \delta(g_i,h_j) 
	\end{eqnarray}
	where $A_{ij}$ is the incidence matrix of the network, $p_{ij}$ is the null model matrix describing the expected probability of interactions between two types of nodes given their degrees, $k_i$ is the degree of resource node $i$ and $d_j$ the degree of consumer node $j$; $g_i$ and $h_i$ are the community indices of nodes $i$ and $j$ and $|{E}|$ is the number of links in the network. After a certain number of extinctions, eventually, bipartite networks break in a set of disconnected components. Thus, to consider the breakup of the network, we introduce the effective modularity $Q_{e}$, which is calculated as the product of the relative size of the largest connected component $S$ and the modularity $Q$: 
	$Q_{e} = SQ$.\\
\begin{figure}[h!]
	\includegraphics[width=0.4\textwidth]{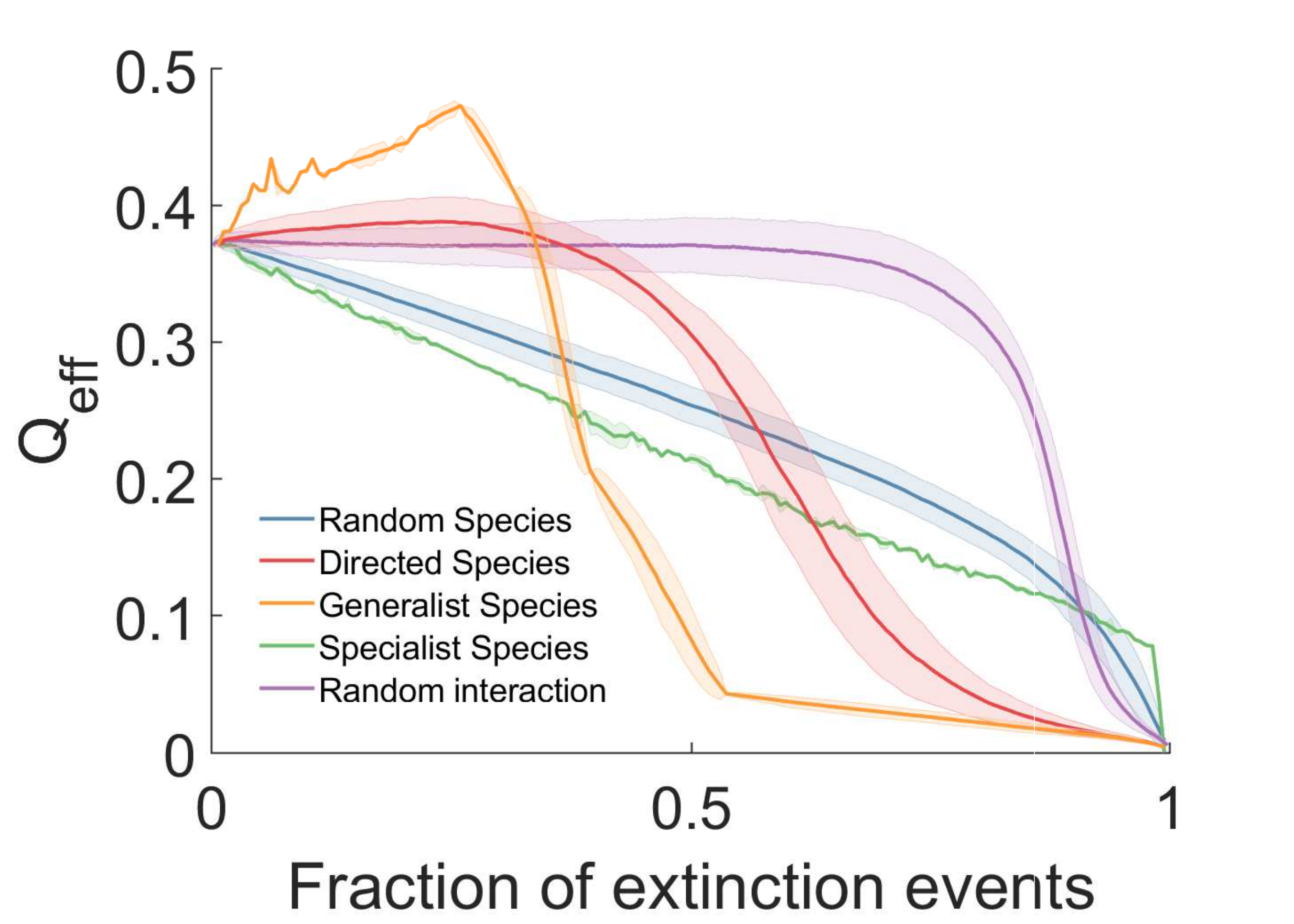}
	\caption{Time evolution of the effective modularity $Q_{e}$. For each realization of the extinction events and for each extinction event, we identify the communities and calculate the measures. Mean values and standard deviations (shaded areas) are obtained after 500 independent realizations for each scenario.}
	\label{Qeff}
\end{figure}
In Fig.~\ref{Qeff} we show the effective modularity for the different extinction scenarios averaged over 500 independent realizations of the dynamics. We observe two behaviors: On the one hand a transition-like behavior, where there exists a critical fraction of extinction events for which the effective modularity sharply decreases, as happens for the generalist, directed and interaction scenarios, with critical fractions of extinction events approximately 0.35, 0.6 and 0.9 respectively. For the generalist scenario, the effective modularity even increases initially. On the other hand, in the random and specialist scenarios, the effective modularity decreases smoothly until the bipartite network is extinct.

\begin{figure}[h!]
	\includegraphics[width=0.5\textwidth]{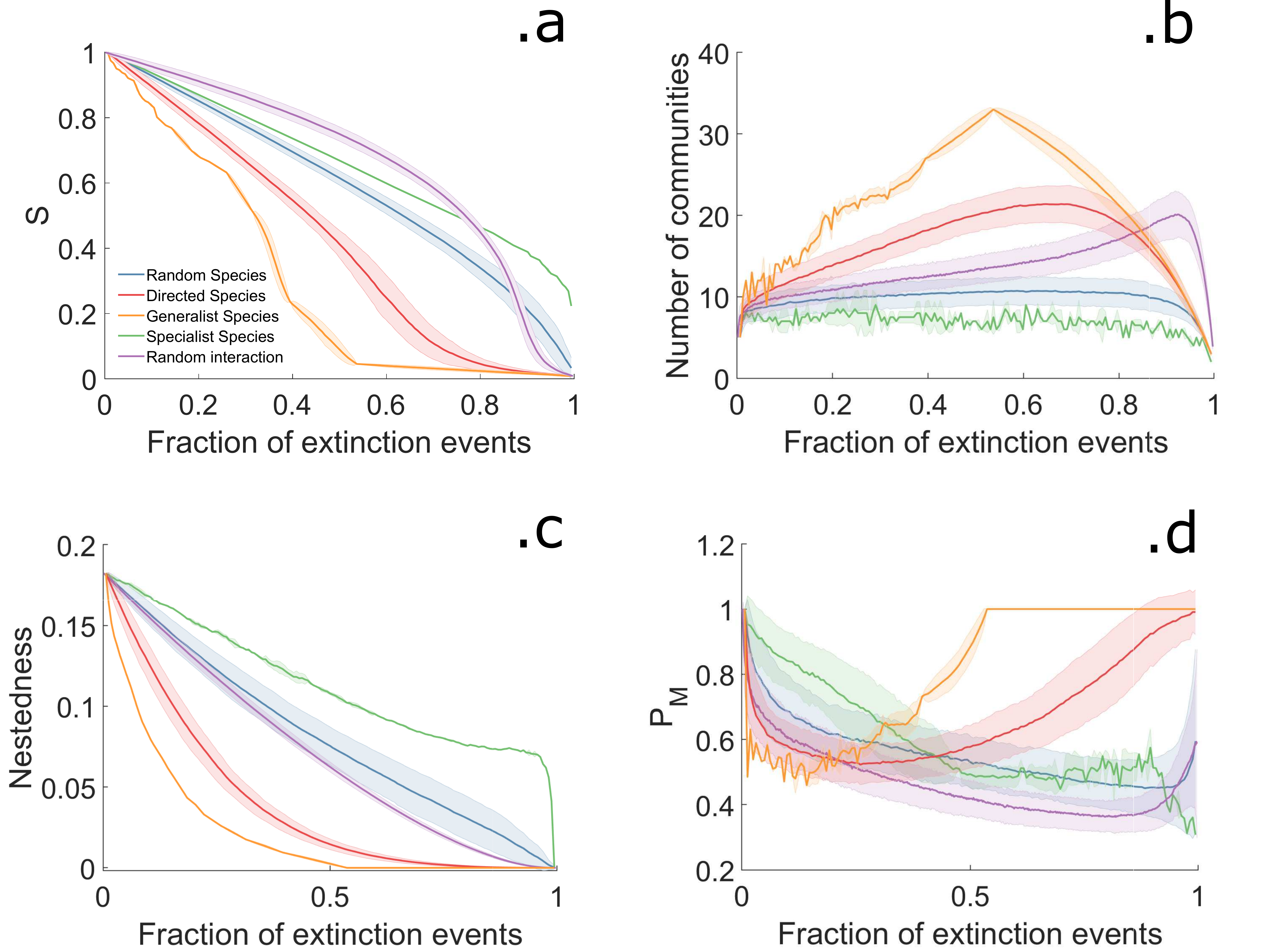}
	\caption{Time evolution of  $(a)$ the size of the largest connected component $S$, $(b)$ the number of communities , $(c)$ nestedness and  $(d)$ persistence $P_{M}$. For each realization of the extinction events and for each extinction event, we identify the communities and calculate the measures. Mean values and standard deviations (shaded areas) are obtained after 500 independent realizations for each scenario.}
	\label{modulePersist}
\end{figure}

In order to gain more insight into the structural reorganization of the network we measure several other quantities as a function of time, including the size of the largest component, the number of communities, the nestedness and the community persistence (Fig.~\ref{modulePersist}). 
Regarding the size of the largest component (Fig.~\ref{modulePersist}.a), which monitors the fragmentation of the network, we observe that the network is more sensitive to generalist species extinctions and collapses faster than the other scenarios, followed closely by the directed extinctions scenario, in line with previous knowledge on robustness under targeted attacks in networks \cite{Albert,Callaway,Cohen,Gallos,Annibale,Huang}. Random node extinctions and specialist extinctions behave very similarly, with a smooth decay of the largest connected component. Last, random interaction extinctions keep the largest connected component larger than in any other scenario until a fraction of extinctions equal to 0.8, where the system rapidly collapses.

The number of communities (Fig.~\ref{modulePersist}.b) is increased initially in all the scenarios, but then remains constant around 10 communities for random node extinctions and specialist extinctions, decaying fast when the fraction of extinction events is almost 1 due to network decomposition. For random interaction extinctions, it behaves similarly in the beginning but toward a fraction of extinctions of 0.6, the number of communities increases rapidly to up to 20 before dropping fast again because of network decomposition. Finally directed and generalist extinctions behave similarly, with a steady increase in the number of communities until a certain fraction of extinction events, where the number of communities decays. This certain fraction is 0.55 and 0.7 for the generalist and directed scenarios respectively and they reach 33 and 21 communities respectively.

The next architectural pattern that we consider here is nestedness, which can be described as the tendency of specialists to interact with proper subsets of the nodes interacting with generalists \cite{Melian,Olesen}. There are several indices for quantifying nestedness depending on whether binary or weighted interaction data are provided. The most commonly used methods are: NTC (Nestedness temperature calculator) \cite{atmar1993}, SR (spectral radius of the adjacency matrix) \cite{staniczenko} and NODF (Nestedness metric based on overlap and decreasing fill) \cite{Almeida}. Here we use the NODF metric to estimate nestedness.

In all the extinction scenarios, nestedness values decrease with extinction events from the very beginning (Fig.~\ref{modulePersist}.c), due to the decrease of the largest degree of the bipartite network, which is positively correlated with NODF \cite{borge2017}. Therefore, the fastest decrease is found for generalist extinctions, followed by directed extinctions. Random node and interaction extinctions behave similarly, decaying more smoothly, almost in a linear fashion, to reach 0 nestedness when $t=1$. Last, the specialist extinctions scenario is the one keeping the network more nested, related to the fact that this scenario is the one diminishing the largest degree the less.

The structural changes in the community structure of the bipartite network can be quantified with the community persistence, {\it i.e.}, the probability that two nodes remain in the same community if they were initially in the same community, $P_{i,j}(M_{i} = M_{j} , t|M_{i} = M_{j} , t_{0})$. We then compute the averages over all node pairs to get the mean persistence $P_{M} = \langle P(M_{i} = M_{j} , t|M_{i} = M_{j} , t_{0})\rangle$. As illustrated in Fig.~\ref{modulePersist}.d, community persistence decays initially fast and as more nodes (or links) are extinct for any scenario. In the random extinction scenario, the persistence decays at a slower rate. In random interaction extinctions, the persistence decays quickly after the extinction of a small fraction of interactions and then still decreases at a lower rate until the extinction of around 90\% of the interactions. We observe an increase of the persistence in the directed and generalist scenarios. This increase is due to the breakup of the bipartite networks where the few interactions remaining corresponds to interactions that originally were identified in the same community.

The variability of community structure is captured with the versatility, $V$. Versatility is a metric of nodal affiliation which describes how closely each node is assigned with a community: $V= 0$ indicates that a node is always assigned to the same community; while $V \gg 0$ determines that it is assigned to different communities depending on the realization \cite{Shinn}. The versatility of a node $j$ is defined as:
\begin{eqnarray}\label{eq}
V(j)=\frac{\sum _{i}\sin (\pi \langle a(i,j)\rangle)}{\sum _{i} \langle a(i,j)\rangle},
\end{eqnarray}
with
\begin{equation*}
a(i,j)=\begin{cases}
1 & \text{if\, $i$\, and\  $j$\, are\, in\, the\, same\, community}.\\
0 & \text{otherwise}.
\end{cases}
\end{equation*}
where $\langle a(i,j)\rangle$ is the expected value of $a(i, j)$ averaged over different realizations evolved to the same fraction of extinction events. A high value of versatility reflects thus a loose community structure, as nodes might be assigned to one or other community, while a low versatility value stands for a robust community structure with well-defined communities. 
\begin{figure}[!h]
	\hspace{-0.7 cm}
	\includegraphics[width=0.5\textwidth]{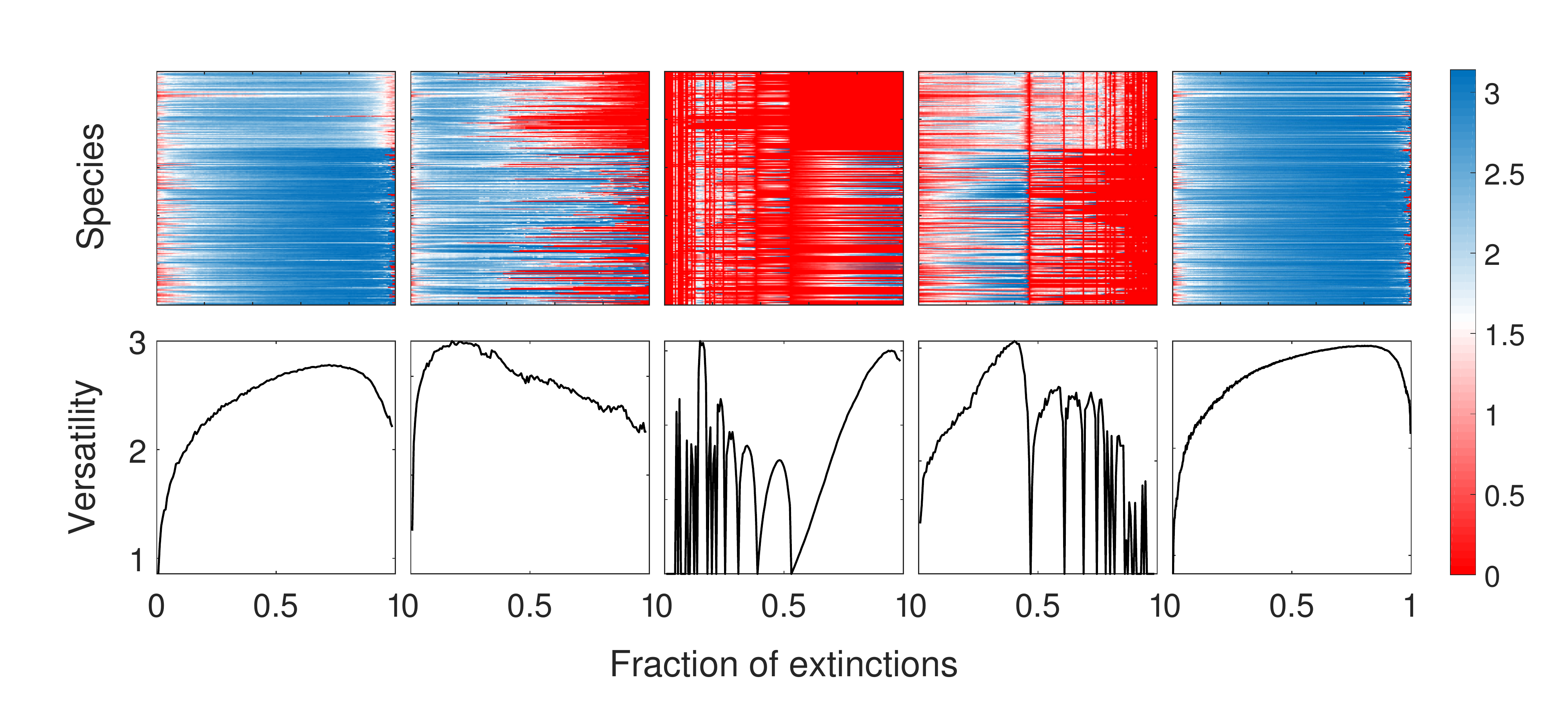}
	\caption{Versatility for the different extinction scenarios. (Top) The versatility defined in
			Eq.~\ref{eq} averaged over 500 realizations is represented in a color scale from $V=0$ (red) to $V=3$ (blue), the x-axis is the fraction of extinction events, and the y-axis represents each species. Each column corresponds the five extinction scenarios: (from left to right) random extinction, directed extinction, generalist extinction, specialist extinction and random interaction extinction. (Bottom) versatility averaged over all species and 500 realizations of the extinction sequences for the different scenarios.}
	\label{SBVersatility}
\end{figure}
For versatility (Fig.~\ref{SBVersatility})  the results show that random node and interaction extinctions behave similarly, with a decreasingly less defined community structure up to 75-80\% of extinction events (growing versatility), followed by a decrease in versatility, associated with a more solid community structure. For directed extinctions, the structure evolves rapidly to a not well-defined community structure (high versatility) and around 25\% of extinction events starts to build a more solid community structure, as versatility decays. For the generalist and specialist extinction scenarios, the picture is a bit more complex. Due to the semi-deterministic nature of the extinction sequences in these scenarios, at some points all of the realizations reach the same configuration, and thus the same community structure, giving rise to 0 versatility. These points are reversed in both scenarios as the sequences are basically reversed. Between these points, the versatility grows because the community structure is less defined as realizations reach different configurations.

\begin{figure}[h!]
	\includegraphics[width=0.45\textwidth]{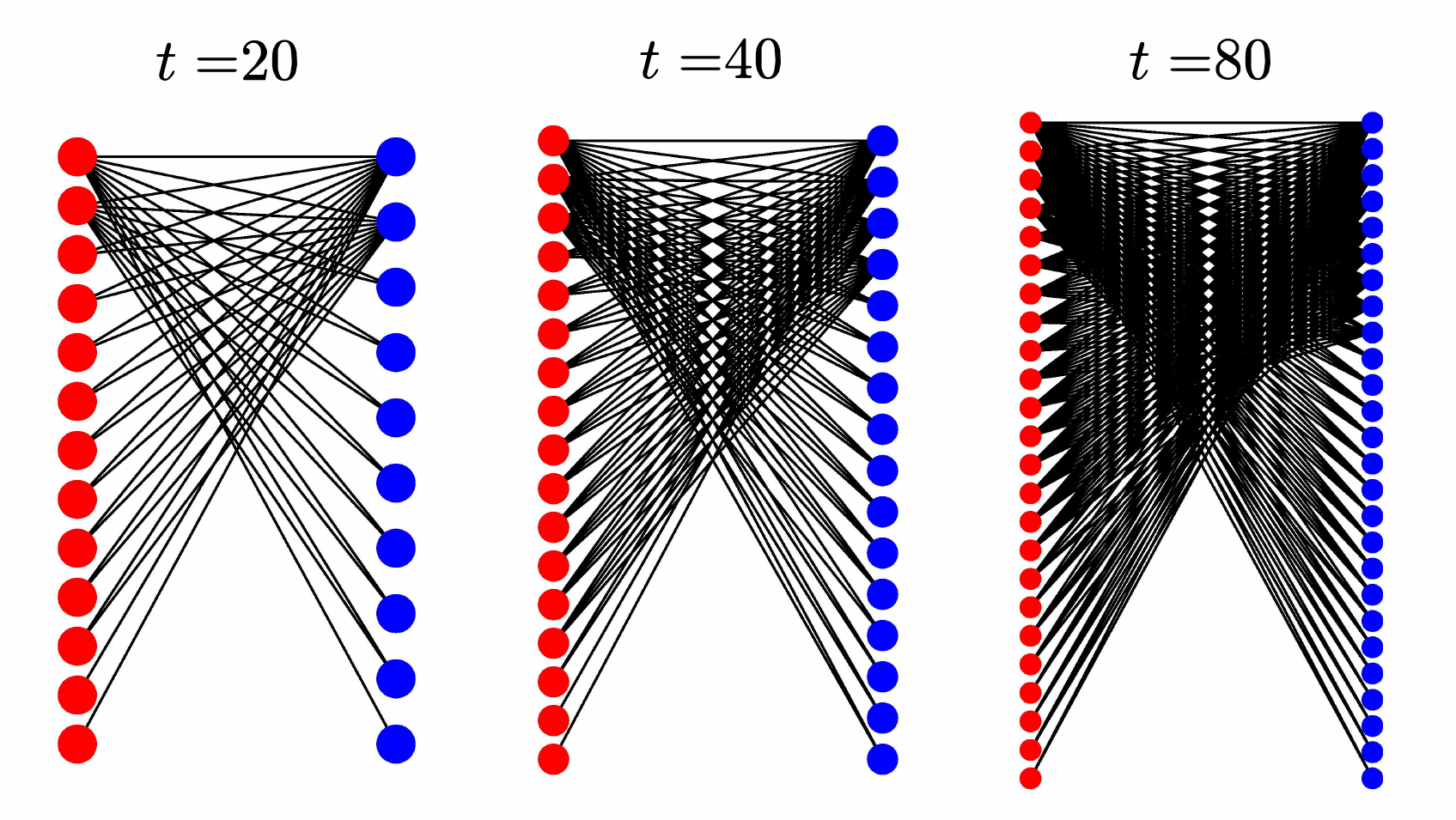}
	\caption{Growth of a bipartite network model.  A realization of the growth of the bipartite network model at iteration time $t = 20, 40, 80$ \cite{Valverde}. Parameters values are: pace of evolutionary change, ($\beta = 10^{-5}$), probability of weight change ($P = 0.1$ ) and link removal threshold ($\theta = 10^{-6}$).}
	\label{netGrowth}
\end{figure}

\section{Structures in growing bipartite Networks}
Additionally, we generate bipartite networks using an evolutionary model of mutualistic webs, through speciation and divergence of weights \cite{Valverde} and then perform numerical simulations to detect the community evolutions (Fig.~\ref{netGrowth}). 

In this model, nodes are considered to be either present or absent, with no role to be played by population size. Some properties such as the heterogeneity in degree distribution or the nestedness of ecological mutualistic networks are captured at the same time by the model.

\begin{figure}[h!]
	\includegraphics[width=0.45\textwidth]{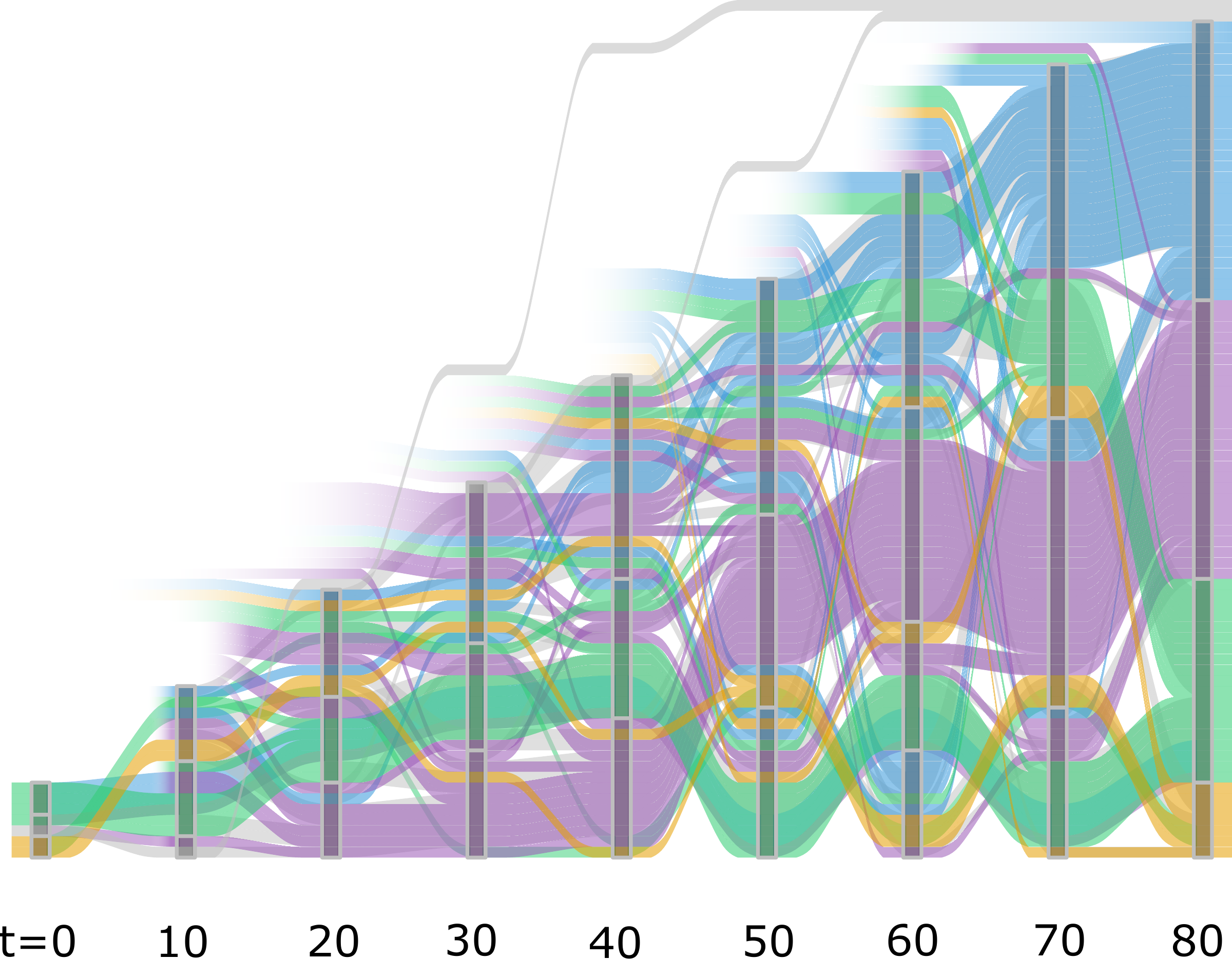}
	\caption{Time evolution of the community structure of a growing network after every 10 time steps for one realization and the same parameters used in Fig.~\ref{netGrowth}.}
	\label{alluvialSpandrel}
\end{figure}

\begin{figure}[h!]
	\includegraphics[width=0.4\textwidth]{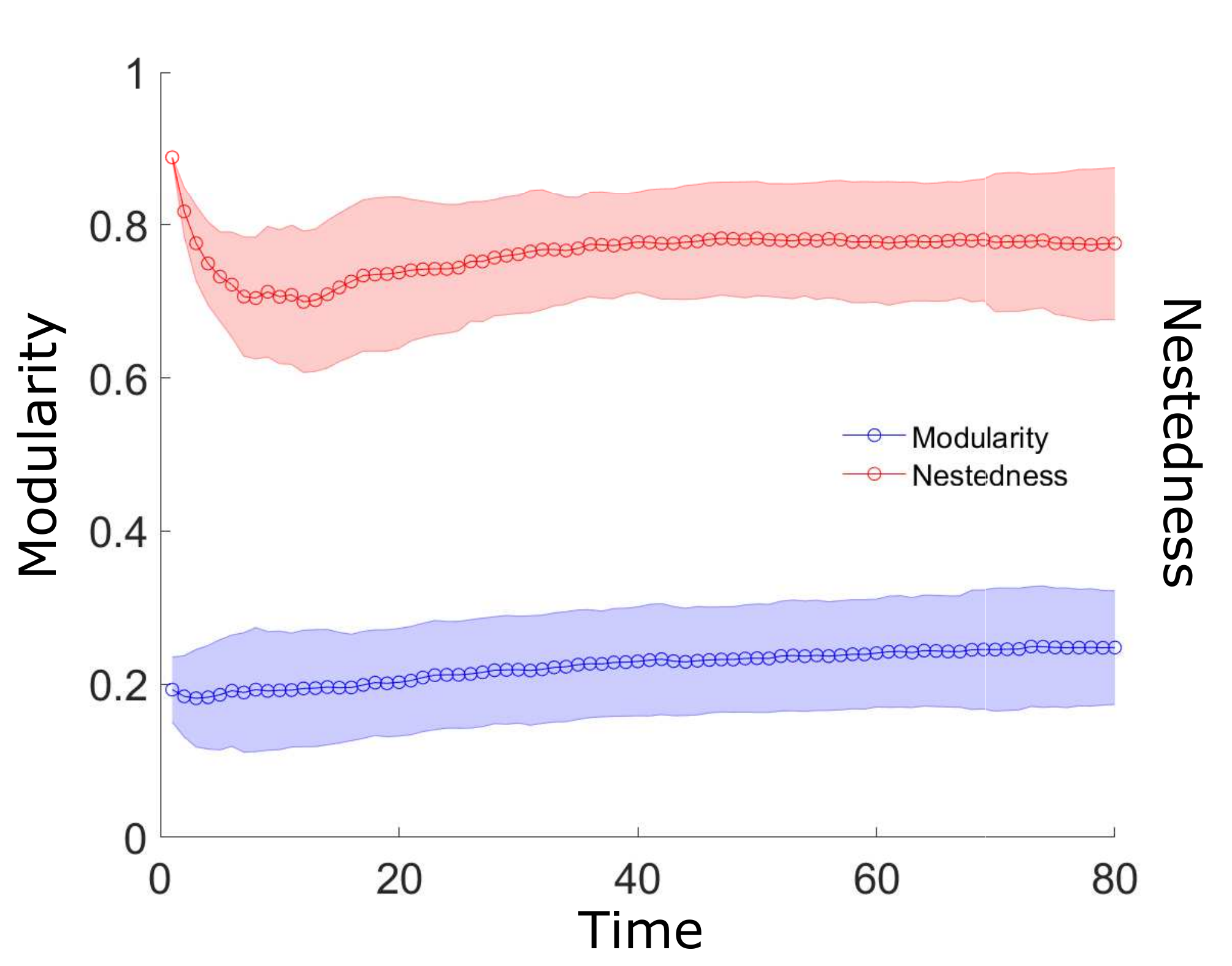}
	\caption{Time evolution of modularity and nestedness. Results are averaged over $1000$ independent realizations of the growing sequences as a function of time. Here we show the mean and standard deviations. The evolutionary growth process keeps the modularity value low whereas nestedness is high. Both remain approximately constant after a very small initial transient.}
	\label{NestMod}
\end{figure}

The community structure is stationary during network growth, with the nodes in each community most probably remaining in the same community (Fig.~\ref{alluvialSpandrel}). The modularity and the nestedness remain low and constant during the network evolution (Fig.~\ref{NestMod}), in contrast with their response when the nodes are removed in any of the scenarios presented above.

\section{Conclusions}

We have studied the evolution of the community structure of an empirical ecological bipartite network in the context of extinction of consumer nodes (in this case pollinators). To do so we have introduced 5 different extinction scenarios to account for different extinction dynamics: 1) random node extinction, 2) directed extinctions, 3) generalist extinctions, 4) specialist extinctions and 5) random interaction extinctions. First, we qualitatively observe that during the initial steps the community structure is not affected too much under random node extinctions and specialist extinctions, in contrast to what happens under generalist, directed and random link extinction mechanisms. We next quantify the changes in the organization of the network and its community structure under the different extinction scenarios with a battery of measures. We show that the community structure is reorganized as a function of the fraction of extinction events, signaled for example by the high versatility values at certain moments of the dynamics. Besides that, and most importantly, our result for the effective modularity shows potential for evaluation of risks of ecosystems, if we know under which kind of extinction dynamics the network is suffering. Considering that modularity is a desirable characteristic in ecological ecosystems, as it can buffer the spread of perturbations \cite{Gardner,Gilarranz}, we can conclude that random node and specialist extinctions are always detrimental for the system, and the response of the system is approximately proportional (in terms of loss of modularity) to the percentage of loss of species.  For random interaction extinctions modularity remains basically unchanged until 90\% of interactions are removed, where the network collapses and modularity suddenly decays. This, therefore, is not such a detrimental extinction dynamics. For directed extinctions, we have also a mostly constant modularity until a drop towards 0 appears when the network fragments. This happens for a fraction of around 0.6 species gone extinct. The problem here is that the response is strongly non-linear and we can go from a relatively high value of the effective modularity to a very low one with just a few species going extinct. This is the same in the case of generalist extinctions, but even more dangerous, as the drop comes at around 35\% of species going extinct. Nevertheless, below that critical value for generalist extinctions the effective modularity actually increases, which may be an effective, but dangerous, way of endowing the network with a with higher modularity -- just removing a few of the most generalist nodes, without taking too many, as we may totally dismantle the network and end up with low modularity.

We postulate, based on our results, that the disappearance of a few generalist species
in mutualistic ecosystems might be a way of protecting the system against the spread of
perturbations, but that this is a dangerous game, given that if too many are gone extinct, the modularity suddenly drops. Ecosystems are typically robust to the removal of a small fraction of the species, extinction of only a single species positioned at the core of the community cause significant to total network collapse \cite{campbell2012}

For a growing bipartite network model, in contrast, the community structure is more stationary than in the extinction scenarios of the real mutualistic networks in the sense that the modularity slightly changes as the network grows in comparison to the variation of the modularity with extinctions.

It rests a challenge to widen the results to the weighted case. The results would depend on whether the weights are distributed homogeneously or heterogeneously. Typically, the weight in mutualistic networks captures the fraction of visits of a pollinator to a plant normalized with the total number of visits. As such, the total weight per plant always adds up to one. If the distribution of weights is homogeneous, our expectation is that the communities would behave similarly. A different case would be for weights distributed mostly along a subset of pollinators. In this case, the results can be greatly affected.


\begin{thebibliography}{33}%
	\makeatletter
	\providecommand \@ifxundefined [1]{%
		\@ifx{#1\undefined}
	}%
	\providecommand \@ifnum [1]{%
		\ifnum #1\expandafter \@firstoftwo
		\else \expandafter \@secondoftwo
		\fi
	}%
	\providecommand \@ifx [1]{%
		\ifx #1\expandafter \@firstoftwo
		\else \expandafter \@secondoftwo
		\fi
	}%
	\providecommand \natexlab [1]{#1}%
	\providecommand \enquote  [1]{``#1''}%
	\providecommand \bibnamefont  [1]{#1}%
	\providecommand \bibfnamefont [1]{#1}%
	\providecommand \citenamefont [1]{#1}%
	\providecommand \href@noop [0]{\@secondoftwo}%
	\providecommand \href [0]{\begingroup \@sanitize@url \@href}%
	\providecommand \@href[1]{\@@startlink{#1}\@@href}%
	\providecommand \@@href[1]{\endgroup#1\@@endlink}%
	\providecommand \@sanitize@url [0]{\catcode `\\12\catcode `\$12\catcode
		`\&12\catcode `\#12\catcode `\^12\catcode `\_12\catcode `\%12\relax}%
	\providecommand \@@startlink[1]{}%
	\providecommand \@@endlink[0]{}%
	\providecommand \url  [0]{\begingroup\@sanitize@url \@url }%
	\providecommand \@url [1]{\endgroup\@href {#1}{\urlprefix }}%
	\providecommand \urlprefix  [0]{URL }%
	\providecommand \Eprint [0]{\href }%
	\providecommand \doibase [0]{http://dx.doi.org/}%
	\providecommand \selectlanguage [0]{\@gobble}%
	\providecommand \bibinfo  [0]{\@secondoftwo}%
	\providecommand \bibfield  [0]{\@secondoftwo}%
	\providecommand \translation [1]{[#1]}%
	\providecommand \BibitemOpen [0]{}%
	\providecommand \bibitemStop [0]{}%
	\providecommand \bibitemNoStop [0]{.\EOS\space}%
	\providecommand \EOS [0]{\spacefactor3000\relax}%
	\providecommand \BibitemShut  [1]{\csname bibitem#1\endcsname}%
	\let\auto@bib@innerbib\@empty
	\bibitem [{\citenamefont {Bascompte}\ \emph
		{et~al.}(2003{\natexlab{a}})\citenamefont {Bascompte}, \citenamefont
		{Jordano}, \citenamefont {Meli{\'a}n},\ and\ \citenamefont
		{Olesen}}]{Jordano}%
	\BibitemOpen
	\bibfield  {author} {\bibinfo {author} {\bibfnamefont {J.}~\bibnamefont
			{Bascompte}}, \bibinfo {author} {\bibfnamefont {P.}~\bibnamefont {Jordano}},
		\bibinfo {author} {\bibfnamefont {C.~J.}\ \bibnamefont {Meli{\'a}n}}, \ and\
		\bibinfo {author} {\bibfnamefont {J.~M.}\ \bibnamefont {Olesen}},\
	}\href@noop {} {\bibfield  {journal} {\bibinfo  {journal} {Proceedings of the
				National Academy of Sciences}\ }\textbf {\bibinfo {volume} {100}},\ \bibinfo
		{pages} {9383} (\bibinfo {year} {2003}{\natexlab{a}})}\BibitemShut {NoStop}%
	\bibitem [{\citenamefont {Gracia-L{\'a}zaro}\ \emph {et~al.}(2018)\citenamefont
		{Gracia-L{\'a}zaro}, \citenamefont {Hern{\'a}ndez}, \citenamefont
		{Borge-Holthoefer},\ and\ \citenamefont {Moreno}}]{Gracia}%
	\BibitemOpen
	\bibfield  {author} {\bibinfo {author} {\bibfnamefont {C.}~\bibnamefont
			{Gracia-L{\'a}zaro}}, \bibinfo {author} {\bibfnamefont {L.}~\bibnamefont
			{Hern{\'a}ndez}}, \bibinfo {author} {\bibfnamefont {J.}~\bibnamefont
			{Borge-Holthoefer}}, \ and\ \bibinfo {author} {\bibfnamefont
			{Y.}~\bibnamefont {Moreno}},\ }\href@noop {} {\bibfield  {journal} {\bibinfo
			{journal} {Scientific reports}\ }\textbf {\bibinfo {volume} {8}},\ \bibinfo
		{pages} {9253} (\bibinfo {year} {2018})}\BibitemShut {NoStop}%
	\bibitem [{\citenamefont {Bascompte}(2010)}]{Bascompte2010}%
	\BibitemOpen
	\bibfield  {author} {\bibinfo {author} {\bibfnamefont {J.}~\bibnamefont
			{Bascompte}},\ }\href@noop {} {\bibfield  {journal} {\bibinfo  {journal}
			{Science (New York, N.Y.)}\ }\textbf {\bibinfo {volume} {329}},\ \bibinfo
		{pages} {765} (\bibinfo {year} {2010})}\BibitemShut {NoStop}%
	\bibitem [{\citenamefont {V{\'a}zquez}\ \emph {et~al.}(2005)\citenamefont
		{V{\'a}zquez}, \citenamefont {Poulin}, \citenamefont {Krasnov},\ and\
		\citenamefont {Shenbrot}}]{Vazquez2005}%
	\BibitemOpen
	\bibfield  {author} {\bibinfo {author} {\bibfnamefont {D.~P.}\ \bibnamefont
			{V{\'a}zquez}}, \bibinfo {author} {\bibfnamefont {R.}~\bibnamefont {Poulin}},
		\bibinfo {author} {\bibfnamefont {B.~R.}\ \bibnamefont {Krasnov}}, \ and\
		\bibinfo {author} {\bibfnamefont {G.~I.}\ \bibnamefont {Shenbrot}},\
	}\href@noop {} {\bibfield  {journal} {\bibinfo  {journal} {Journal of Animal
				Ecology}\ }\textbf {\bibinfo {volume} {74}},\ \bibinfo {pages} {946}
		(\bibinfo {year} {2005})}\BibitemShut {NoStop}%
	\bibitem [{\citenamefont {May}(1972)}]{May1972}%
	\BibitemOpen
	\bibfield  {author} {\bibinfo {author} {\bibfnamefont {R.~M.}\ \bibnamefont
			{May}},\ }\href@noop {} {\bibfield  {journal} {\bibinfo  {journal} {Nature}\
		}\textbf {\bibinfo {volume} {238}},\ \bibinfo {pages} {413} (\bibinfo {year}
		{1972})}\BibitemShut {NoStop}%
	\bibitem [{\citenamefont {Burgos}\ \emph {et~al.}(2007)\citenamefont {Burgos},
		\citenamefont {Ceva}, \citenamefont {Perazzo}, \citenamefont {Devoto},
		\citenamefont {Medan}, \citenamefont {Zimmermann},\ and\ \citenamefont
		{Delbue}}]{Burgos2007}%
	\BibitemOpen
	\bibfield  {author} {\bibinfo {author} {\bibfnamefont {E.}~\bibnamefont
			{Burgos}}, \bibinfo {author} {\bibfnamefont {H.}~\bibnamefont {Ceva}},
		\bibinfo {author} {\bibfnamefont {R.~P.}\ \bibnamefont {Perazzo}}, \bibinfo
		{author} {\bibfnamefont {M.}~\bibnamefont {Devoto}}, \bibinfo {author}
		{\bibfnamefont {D.}~\bibnamefont {Medan}}, \bibinfo {author} {\bibfnamefont
			{M.}~\bibnamefont {Zimmermann}}, \ and\ \bibinfo {author} {\bibfnamefont
			{A.~M.}\ \bibnamefont {Delbue}},\ }\href@noop {} {\bibfield  {journal}
		{\bibinfo  {journal} {Journal of theoretical biology}\ }\textbf {\bibinfo
			{volume} {249}},\ \bibinfo {pages} {307} (\bibinfo {year}
		{2007})}\BibitemShut {NoStop}%
	\bibitem [{\citenamefont {Staniczenko}\ \emph {et~al.}(2010)\citenamefont
		{Staniczenko}, \citenamefont {Lewis}, \citenamefont {Jones},\ and\
		\citenamefont {Reed-Tsochas}}]{Staniczenko2010}%
	\BibitemOpen
	\bibfield  {author} {\bibinfo {author} {\bibfnamefont {P.~P.}\ \bibnamefont
			{Staniczenko}}, \bibinfo {author} {\bibfnamefont {O.~T.}\ \bibnamefont
			{Lewis}}, \bibinfo {author} {\bibfnamefont {N.~S.}\ \bibnamefont {Jones}}, \
		and\ \bibinfo {author} {\bibfnamefont {F.}~\bibnamefont {Reed-Tsochas}},\
	}\href@noop {} {\bibfield  {journal} {\bibinfo  {journal} {Ecology letters}\
		}\textbf {\bibinfo {volume} {13}},\ \bibinfo {pages} {891} (\bibinfo {year}
		{2010})}\BibitemShut {NoStop}%
	\bibitem [{\citenamefont {Gardner}\ and\ \citenamefont
		{Ashby}(1970)}]{Gardner}%
	\BibitemOpen
	\bibfield  {author} {\bibinfo {author} {\bibfnamefont {M.~R.}\ \bibnamefont
			{Gardner}}\ and\ \bibinfo {author} {\bibfnamefont {W.~R.}\ \bibnamefont
			{Ashby}},\ }\href@noop {} {\bibfield  {journal} {\bibinfo  {journal}
			{Nature}\ }\textbf {\bibinfo {volume} {228}},\ \bibinfo {pages} {784}
		(\bibinfo {year} {1970})}\BibitemShut {NoStop}%
	\bibitem [{\citenamefont {{J. Gilarranz}}\ \emph {et~al.}(2017)\citenamefont
		{{J. Gilarranz}}, \citenamefont {Rayfield}, \citenamefont
		{Li{\~{n}}an-Cembrano}, \citenamefont {Bascompte},\ and\ \citenamefont
		{Gonzalez}}]{Gilarranz}%
	\BibitemOpen
	\bibfield  {author} {\bibinfo {author} {\bibfnamefont {L.}~\bibnamefont {{J.
					Gilarranz}}}, \bibinfo {author} {\bibfnamefont {B.}~\bibnamefont {Rayfield}},
		\bibinfo {author} {\bibfnamefont {G.}~\bibnamefont {Li{\~{n}}an-Cembrano}},
		\bibinfo {author} {\bibfnamefont {J.}~\bibnamefont {Bascompte}}, \ and\
		\bibinfo {author} {\bibfnamefont {A.}~\bibnamefont {Gonzalez}},\ }\href@noop
	{} {\emph {\bibinfo {title} {Science}}},\ Vol.\ \bibinfo {volume} {357}\
	(\bibinfo {year} {2017})\ p.\ \bibinfo {pages} {199}\BibitemShut {NoStop}%
	\bibitem [{\citenamefont {Jane}\ \emph {et~al.}(2004)\citenamefont {Jane},
		\citenamefont {M.},\ and\ \citenamefont {V.}}]{Memmott}%
	\BibitemOpen
	\bibfield  {author} {\bibinfo {author} {\bibfnamefont {M.}~\bibnamefont
			{Jane}}, \bibinfo {author} {\bibfnamefont {W.~N.}\ \bibnamefont {M.}}, \ and\
		\bibinfo {author} {\bibfnamefont {P.~M.}\ \bibnamefont {V.}},\ }\href
	{https://doi.org/10.1098/rspb.2004.2909} {\bibfield  {journal} {\bibinfo
			{journal} {Proceedings of the Royal Society of London. Series B: Biological
				Sciences}\ }\textbf {\bibinfo {volume} {271}},\ \bibinfo {pages} {2605}
		(\bibinfo {year} {2004})}\BibitemShut {NoStop}%
	\bibitem [{\citenamefont {Potts}\ \emph {et~al.}(2010)\citenamefont {Potts},
		\citenamefont {Biesmeijer}, \citenamefont {Kremen}, \citenamefont {Neumann},
		\citenamefont {Schweiger},\ and\ \citenamefont {Kunin}}]{POTTS}%
	\BibitemOpen
	\bibfield  {author} {\bibinfo {author} {\bibfnamefont {S.~G.}\ \bibnamefont
			{Potts}}, \bibinfo {author} {\bibfnamefont {J.~C.}\ \bibnamefont
			{Biesmeijer}}, \bibinfo {author} {\bibfnamefont {C.}~\bibnamefont {Kremen}},
		\bibinfo {author} {\bibfnamefont {P.}~\bibnamefont {Neumann}}, \bibinfo
		{author} {\bibfnamefont {O.}~\bibnamefont {Schweiger}}, \ and\ \bibinfo
		{author} {\bibfnamefont {W.~E.}\ \bibnamefont {Kunin}},\ }\href
	{http://www.sciencedirect.com/science/article/pii/S0169534710000364}
	{\bibfield  {journal} {\bibinfo  {journal} {Trends in Ecology \& Evolution}\
		}\textbf {\bibinfo {volume} {25}},\ \bibinfo {pages} {345 } (\bibinfo {year}
		{2010})}\BibitemShut {NoStop}%
	\bibitem [{\citenamefont {Jordano}\ \emph {et~al.}(2006)\citenamefont
		{Jordano}, \citenamefont {Bascompte},\ and\ \citenamefont {Olesen}}]{Pedro}%
	\BibitemOpen
	\bibfield  {author} {\bibinfo {author} {\bibfnamefont {P.}~\bibnamefont
			{Jordano}}, \bibinfo {author} {\bibfnamefont {J.}~\bibnamefont {Bascompte}},
		\ and\ \bibinfo {author} {\bibfnamefont {J.~M.}\ \bibnamefont {Olesen}},\
	}in\ \href@noop {} {\emph {\bibinfo {booktitle} {Plant-pollinator
				interactions: from specialization to generalization}}}\ (\bibinfo
	{publisher} {University of Chicago Press},\ \bibinfo {address} {Chicago},\
	\bibinfo {year} {2006})\ pp.\ \bibinfo {pages} {173--199}\BibitemShut
	{NoStop}%
	\bibitem [{\citenamefont {Valverde}\ \emph {et~al.}(2018)\citenamefont
		{Valverde}, \citenamefont {Pi{\~{n}}ero}, \citenamefont {Corominas-Murtra},
		\citenamefont {Montoya}, \citenamefont {Joppa},\ and\ \citenamefont
		{Sol{\'{e}}}}]{Valverde}%
	\BibitemOpen
	\bibfield  {author} {\bibinfo {author} {\bibfnamefont {S.}~\bibnamefont
			{Valverde}}, \bibinfo {author} {\bibfnamefont {J.}~\bibnamefont
			{Pi{\~{n}}ero}}, \bibinfo {author} {\bibfnamefont {B.}~\bibnamefont
			{Corominas-Murtra}}, \bibinfo {author} {\bibfnamefont {J.}~\bibnamefont
			{Montoya}}, \bibinfo {author} {\bibfnamefont {L.}~\bibnamefont {Joppa}}, \
		and\ \bibinfo {author} {\bibfnamefont {R.}~\bibnamefont {Sol{\'{e}}}},\
	}\href {https://www.ncbi.nlm.nih.gov/pubmed/29158553
		https://www.ncbi.nlm.nih.gov/pmc/PMC6025779/} {\bibfield  {journal} {\bibinfo
			{journal} {Nature ecology {\&} evolution}\ }\textbf {\bibinfo {volume}
			{2}},\ \bibinfo {pages} {94} (\bibinfo {year} {2018})}\BibitemShut {NoStop}%
	\bibitem [{\citenamefont {Traveset}\ \emph {et~al.}(2017)\citenamefont
		{Traveset}, \citenamefont {Tur},\ and\ \citenamefont
		{Egu{\'{i}}luz}}]{Traveset}%
	\BibitemOpen
	\bibfield  {author} {\bibinfo {author} {\bibfnamefont {A.}~\bibnamefont
			{Traveset}}, \bibinfo {author} {\bibfnamefont {C.}~\bibnamefont {Tur}}, \
		and\ \bibinfo {author} {\bibfnamefont {V.~M.}\ \bibnamefont
			{Egu{\'{i}}luz}},\ }\href {https://doi.org/10.1038/s41598-017-07037-7}
	{\bibfield  {journal} {\bibinfo  {journal} {Scientific Reports}\ }\textbf
		{\bibinfo {volume} {7}},\ \bibinfo {pages} {6915} (\bibinfo {year}
		{2017})}\BibitemShut {NoStop}%
	\bibitem [{\citenamefont {Albert}\ and\ \citenamefont
		{Barabأ،si}(2000)}]{Albert}%
	\BibitemOpen
	\bibfield  {author} {\bibinfo {author} {\bibfnamefont {H.~J.}\ \bibnamefont
			{Albert}, \bibfnamefont {Réka}}\ and\ \bibinfo {author} {\bibfnamefont
			{A.-L.}\ \bibnamefont {Barabási}},\ }\href@noop {} {\bibfield  {journal}
		{\bibinfo  {journal} {Nature}\ }\textbf {\bibinfo {volume} {406}},\ \bibinfo
		{pages} {378} (\bibinfo {year} {2000})}\BibitemShut {NoStop}%
	\bibitem [{\citenamefont {Evans}\ \emph {et~al.}(2013)\citenamefont {Evans},
		\citenamefont {Pocock},\ and\ \citenamefont {Memmott}}]{Evans}%
	\BibitemOpen
	\bibfield  {author} {\bibinfo {author} {\bibfnamefont {D.~M.}\ \bibnamefont
			{Evans}}, \bibinfo {author} {\bibfnamefont {M.~J.~O.}\ \bibnamefont
			{Pocock}}, \ and\ \bibinfo {author} {\bibfnamefont {J.}~\bibnamefont
			{Memmott}},\ }\href
	{https://onlinelibrary.wiley.com/doi/abs/10.1111/ele.12117} {\bibfield
		{journal} {\bibinfo  {journal} {Ecology Letters}\ }\textbf {\bibinfo {volume}
			{16}},\ \bibinfo {pages} {844} (\bibinfo {year} {2013})}\BibitemShut
	{NoStop}%
	\bibitem [{\citenamefont {Gao}\ \emph {et~al.}(2016)\citenamefont {Gao},
		\citenamefont {Barzel},\ and\ \citenamefont {Barabasi}}]{Gao}%
	\BibitemOpen
	\bibfield  {author} {\bibinfo {author} {\bibfnamefont {J.}~\bibnamefont
			{Gao}}, \bibinfo {author} {\bibfnamefont {B.}~\bibnamefont {Barzel}}, \ and\
		\bibinfo {author} {\bibfnamefont {A.-L.}\ \bibnamefont {Barabasi}},\
	}\href@noop {} {\bibfield  {journal} {\bibinfo  {journal} {Nature}\ }\textbf
		{\bibinfo {volume} {530}},\ \bibinfo {pages} {307} (\bibinfo {year}
		{2016})}\BibitemShut {NoStop}%
	\bibitem [{\citenamefont {Vأ،zquez}\ and\ \citenamefont {Aizen}(2003)}]{Aizen}%
	\BibitemOpen
	\bibfield  {author} {\bibinfo {author} {\bibfnamefont {D.}~\bibnamefont
			{Vázquez}}\ and\ \bibinfo {author} {\bibfnamefont {M.}~\bibnamefont
			{Aizen}},\ }\href@noop {} {\bibfield  {journal} {\bibinfo  {journal}
			{Ecology}\ }\textbf {\bibinfo {volume} {84}},\ \bibinfo {pages} {2493}
		(\bibinfo {year} {2003})}\BibitemShut {NoStop}%
	\bibitem [{\citenamefont {Flores}\ \emph {et~al.}()\citenamefont {Flores},
		\citenamefont {Poisot}, \citenamefont {Valverde},\ and\ \citenamefont
		{Weitz}}]{bimat}%
	\BibitemOpen
	\bibfield  {author} {\bibinfo {author} {\bibfnamefont {C.~O.}\ \bibnamefont
			{Flores}}, \bibinfo {author} {\bibfnamefont {T.}~\bibnamefont {Poisot}},
		\bibinfo {author} {\bibfnamefont {S.}~\bibnamefont {Valverde}}, \ and\
		\bibinfo {author} {\bibfnamefont {J.~S.}\ \bibnamefont {Weitz}},\ }\href
	{https://besjournals.onlinelibrary.wiley.com/doi/abs/10.1111/2041-210X.12458}
	{\bibfield  {journal} {\bibinfo  {journal} {Methods in Ecology and
				Evolution}\ }\textbf {\bibinfo {volume} {7}},\ \bibinfo {pages}
		{127}}\BibitemShut {NoStop}%
	\bibitem [{\citenamefont {Barber}(2007)}]{Barber}%
	\BibitemOpen
	\bibfield  {author} {\bibinfo {author} {\bibfnamefont {M.~J.}\ \bibnamefont
			{Barber}},\ }\href {https://link.aps.org/doi/10.1103/PhysRevE.76.066102}
	{\bibfield  {journal} {\bibinfo  {journal} {Phys. Rev. E}\ }\textbf {\bibinfo
			{volume} {76}},\ \bibinfo {pages} {066102} (\bibinfo {year}
		{2007})}\BibitemShut {NoStop}%
	\bibitem [{\citenamefont {Callaway}\ \emph {et~al.}(2000)\citenamefont
		{Callaway}, \citenamefont {Newman}, \citenamefont {Strogatz},\ and\
		\citenamefont {Watts}}]{Callaway}%
	\BibitemOpen
	\bibfield  {author} {\bibinfo {author} {\bibfnamefont {D.~S.}\ \bibnamefont
			{Callaway}}, \bibinfo {author} {\bibfnamefont {M.~E.~J.}\ \bibnamefont
			{Newman}}, \bibinfo {author} {\bibfnamefont {S.~H.}\ \bibnamefont
			{Strogatz}}, \ and\ \bibinfo {author} {\bibfnamefont {D.~J.}\ \bibnamefont
			{Watts}},\ }\href {https://link.aps.org/doi/10.1103/PhysRevLett.85.5468}
	{\bibfield  {journal} {\bibinfo  {journal} {Phys. Rev. Lett.}\ }\textbf
		{\bibinfo {volume} {85}},\ \bibinfo {pages} {5468} (\bibinfo {year}
		{2000})}\BibitemShut {NoStop}%
	\bibitem [{\citenamefont {Cohen}\ \emph {et~al.}(2001)\citenamefont {Cohen},
		\citenamefont {Erez}, \citenamefont {ben Avraham},\ and\ \citenamefont
		{Havlin}}]{Cohen}%
	\BibitemOpen
	\bibfield  {author} {\bibinfo {author} {\bibfnamefont {R.}~\bibnamefont
			{Cohen}}, \bibinfo {author} {\bibfnamefont {K.}~\bibnamefont {Erez}},
		\bibinfo {author} {\bibfnamefont {D.}~\bibnamefont {ben Avraham}}, \ and\
		\bibinfo {author} {\bibfnamefont {S.}~\bibnamefont {Havlin}},\ }\href
	{https://link.aps.org/doi/10.1103/PhysRevLett.86.3682} {\bibfield  {journal}
		{\bibinfo  {journal} {Phys. Rev. Lett.}\ }\textbf {\bibinfo {volume} {86}},\
		\bibinfo {pages} {3682} (\bibinfo {year} {2001})}\BibitemShut {NoStop}%
	\bibitem [{\citenamefont {Gallos}\ \emph {et~al.}(2005)\citenamefont {Gallos},
		\citenamefont {Cohen}, \citenamefont {Argyrakis}, \citenamefont {Bunde},\
		and\ \citenamefont {Havlin}}]{Gallos}%
	\BibitemOpen
	\bibfield  {author} {\bibinfo {author} {\bibfnamefont {L.~K.}\ \bibnamefont
			{Gallos}}, \bibinfo {author} {\bibfnamefont {R.}~\bibnamefont {Cohen}},
		\bibinfo {author} {\bibfnamefont {P.}~\bibnamefont {Argyrakis}}, \bibinfo
		{author} {\bibfnamefont {A.}~\bibnamefont {Bunde}}, \ and\ \bibinfo {author}
		{\bibfnamefont {S.}~\bibnamefont {Havlin}},\ }\href
	{https://link.aps.org/doi/10.1103/PhysRevLett.94.188701} {\bibfield
		{journal} {\bibinfo  {journal} {Phys. Rev. Lett.}\ }\textbf {\bibinfo
			{volume} {94}},\ \bibinfo {pages} {188701} (\bibinfo {year}
		{2005})}\BibitemShut {NoStop}%
	\bibitem [{\citenamefont {Annibale}\ \emph {et~al.}(2010)\citenamefont
		{Annibale}, \citenamefont {Coolen},\ and\ \citenamefont
		{Bianconi}}]{Annibale}%
	\BibitemOpen
	\bibfield  {author} {\bibinfo {author} {\bibfnamefont {A.}~\bibnamefont
			{Annibale}}, \bibinfo {author} {\bibfnamefont {A.}~\bibnamefont {Coolen}}, \
		and\ \bibinfo {author} {\bibfnamefont {G.}~\bibnamefont {Bianconi}},\
	}\href@noop {} {\bibfield  {journal} {\bibinfo  {journal} {Journal of Physics
				A: Mathematical and Theoretical}\ }\textbf {\bibinfo {volume} {43}},\
		\bibinfo {pages} {395001} (\bibinfo {year} {2010})}\BibitemShut {NoStop}%
	\bibitem [{\citenamefont {Huang}\ \emph {et~al.}(2011)\citenamefont {Huang},
		\citenamefont {Gao}, \citenamefont {Buldyrev}, \citenamefont {Havlin},\ and\
		\citenamefont {Stanley}}]{Huang}%
	\BibitemOpen
	\bibfield  {author} {\bibinfo {author} {\bibfnamefont {X.}~\bibnamefont
			{Huang}}, \bibinfo {author} {\bibfnamefont {J.}~\bibnamefont {Gao}}, \bibinfo
		{author} {\bibfnamefont {S.~V.}\ \bibnamefont {Buldyrev}}, \bibinfo {author}
		{\bibfnamefont {S.}~\bibnamefont {Havlin}}, \ and\ \bibinfo {author}
		{\bibfnamefont {H.~E.}\ \bibnamefont {Stanley}},\ }\href
	{https://link.aps.org/doi/10.1103/PhysRevE.83.065101} {\bibfield  {journal}
		{\bibinfo  {journal} {Phys. Rev. E}\ }\textbf {\bibinfo {volume} {83}},\
		\bibinfo {pages} {065101} (\bibinfo {year} {2011})}\BibitemShut {NoStop}%
	\bibitem [{\citenamefont {Bascompte}\ \emph
		{et~al.}(2003{\natexlab{b}})\citenamefont {Bascompte}, \citenamefont
		{Jordano}, \citenamefont {Melian},\ and\ \citenamefont {Olesen}}]{Melian}%
	\BibitemOpen
	\bibfield  {author} {\bibinfo {author} {\bibfnamefont {J.}~\bibnamefont
			{Bascompte}}, \bibinfo {author} {\bibfnamefont {P.}~\bibnamefont {Jordano}},
		\bibinfo {author} {\bibfnamefont {C.~J.}\ \bibnamefont {Melian}}, \ and\
		\bibinfo {author} {\bibfnamefont {J.~M.}\ \bibnamefont {Olesen}},\ }\href
	{http://www.pnas.org/cgi/doi/10.1073/pnas.1633576100} {\bibfield  {journal}
		{\bibinfo  {journal} {Proceedings of the National Academy of Sciences}\
		}\textbf {\bibinfo {volume} {100}},\ \bibinfo {pages} {9383} (\bibinfo {year}
		{2003}{\natexlab{b}})}\BibitemShut {NoStop}%
	\bibitem [{\citenamefont {Jordano}\ \emph {et~al.}(2003)\citenamefont
		{Jordano}, \citenamefont {Bascompte},\ and\ \citenamefont {Olesen}}]{Olesen}%
	\BibitemOpen
	\bibfield  {author} {\bibinfo {author} {\bibfnamefont {P.}~\bibnamefont
			{Jordano}}, \bibinfo {author} {\bibfnamefont {J.}~\bibnamefont {Bascompte}},
		\ and\ \bibinfo {author} {\bibfnamefont {J.~M.}\ \bibnamefont {Olesen}},\
	}\href
	{https://onlinelibrary.wiley.com/doi/abs/10.1046/j.1461-0248.2003.00403.x}
	{\bibfield  {journal} {\bibinfo  {journal} {Ecology Letters}\ }\textbf
		{\bibinfo {volume} {6}},\ \bibinfo {pages} {69} (\bibinfo {year}
		{2003})}\BibitemShut {NoStop}%
	\bibitem [{\citenamefont {Atmar}\ and\ \citenamefont
		{Patterson}(1993)}]{atmar1993}%
	\BibitemOpen
	\bibfield  {author} {\bibinfo {author} {\bibfnamefont {W.}~\bibnamefont
			{Atmar}}\ and\ \bibinfo {author} {\bibfnamefont {B.~D.}\ \bibnamefont
			{Patterson}},\ }\href@noop {} {\bibfield  {journal} {\bibinfo  {journal}
			{Oecologia}\ }\textbf {\bibinfo {volume} {96}},\ \bibinfo {pages} {373}
		(\bibinfo {year} {1993})}\BibitemShut {NoStop}%
	\bibitem [{\citenamefont {Staniczenko}\ \emph {et~al.}(2013)\citenamefont
		{Staniczenko}, \citenamefont {Kopp},\ and\ \citenamefont
		{Allesina}}]{staniczenko}%
	\BibitemOpen
	\bibfield  {author} {\bibinfo {author} {\bibfnamefont {P.~P.}\ \bibnamefont
			{Staniczenko}}, \bibinfo {author} {\bibfnamefont {J.~C.}\ \bibnamefont
			{Kopp}}, \ and\ \bibinfo {author} {\bibfnamefont {S.}~\bibnamefont
			{Allesina}},\ }\href@noop {} {\bibfield  {journal} {\bibinfo  {journal}
			{Nature communications}\ }\textbf {\bibinfo {volume} {4}},\ \bibinfo {pages}
		{1391} (\bibinfo {year} {2013})}\BibitemShut {NoStop}%
	\bibitem [{\citenamefont {Almeida-Neto}\ \emph {et~al.}(2008)\citenamefont
		{Almeida-Neto}, \citenamefont {Guimarأ£es}, \citenamefont {Guimarأ£es~Jr},
		\citenamefont {Loyola},\ and\ \citenamefont {Ulrich}}]{Almeida}%
	\BibitemOpen
	\bibfield  {author} {\bibinfo {author} {\bibfnamefont {M.}~\bibnamefont
			{Almeida-Neto}}, \bibinfo {author} {\bibfnamefont {P.}~\bibnamefont
			{Guimarães}}, \bibinfo {author} {\bibfnamefont {P.~R.}\ \bibnamefont
			{Guimarães~Jr}}, \bibinfo {author} {\bibfnamefont {R.~D.}\ \bibnamefont
			{Loyola}}, \ and\ \bibinfo {author} {\bibfnamefont {W.}~\bibnamefont
			{Ulrich}},\ }\href@noop {} {\bibfield  {journal} {\bibinfo  {journal}
			{Oikos}\ }\textbf {\bibinfo {volume} {117}},\ \bibinfo {pages} {1227}
		(\bibinfo {year} {2008})}\BibitemShut {NoStop}%
	\bibitem [{\citenamefont {Borge-Holthoefer}\ \emph {et~al.}(2017)\citenamefont
		{Borge-Holthoefer}, \citenamefont {Ba{\~n}os}, \citenamefont
		{Gracia-L{\'a}zaro},\ and\ \citenamefont {Moreno}}]{borge2017}%
	\BibitemOpen
	\bibfield  {author} {\bibinfo {author} {\bibfnamefont {J.}~\bibnamefont
			{Borge-Holthoefer}}, \bibinfo {author} {\bibfnamefont {R.~A.}\ \bibnamefont
			{Ba{\~n}os}}, \bibinfo {author} {\bibfnamefont {C.}~\bibnamefont
			{Gracia-L{\'a}zaro}}, \ and\ \bibinfo {author} {\bibfnamefont
			{Y.}~\bibnamefont {Moreno}},\ }\href@noop {} {\bibfield  {journal} {\bibinfo
			{journal} {Scientific reports}\ }\textbf {\bibinfo {volume} {7}},\ \bibinfo
		{pages} {41673} (\bibinfo {year} {2017})}\BibitemShut {NoStop}%
	\bibitem [{\citenamefont {Shinn}\ \emph {et~al.}(2017)\citenamefont {Shinn},
		\citenamefont {Romero-Garcia}, \citenamefont {Seidlitz}, \citenamefont
		{V{\'{a}}{\v{s}}a}, \citenamefont {V{\'{e}}rtes},\ and\ \citenamefont
		{Bullmore}}]{Shinn}%
	\BibitemOpen
	\bibfield  {author} {\bibinfo {author} {\bibfnamefont {M.}~\bibnamefont
			{Shinn}}, \bibinfo {author} {\bibfnamefont {R.}~\bibnamefont
			{Romero-Garcia}}, \bibinfo {author} {\bibfnamefont {J.}~\bibnamefont
			{Seidlitz}}, \bibinfo {author} {\bibfnamefont {F.}~\bibnamefont
			{V{\'{a}}{\v{s}}a}}, \bibinfo {author} {\bibfnamefont {P.~E.}\ \bibnamefont
			{V{\'{e}}rtes}}, \ and\ \bibinfo {author} {\bibfnamefont {E.}~\bibnamefont
			{Bullmore}},\ }\href {https://doi.org/10.1038/s41598-017-03394-5} {\bibfield
		{journal} {\bibinfo  {journal} {Scientific Reports}\ }\textbf {\bibinfo
			{volume} {7}},\ \bibinfo {pages} {4273} (\bibinfo {year} {2017})}\BibitemShut
	{NoStop}%
	\bibitem [{\citenamefont {Campbell}\ \emph {et~al.}(2012)\citenamefont
		{Campbell}, \citenamefont {Yang}, \citenamefont {Shea},\ and\ \citenamefont
		{Albert}}]{campbell2012}%
	\BibitemOpen
	\bibfield  {author} {\bibinfo {author} {\bibfnamefont {C.}~\bibnamefont
			{Campbell}}, \bibinfo {author} {\bibfnamefont {S.}~\bibnamefont {Yang}},
		\bibinfo {author} {\bibfnamefont {K.}~\bibnamefont {Shea}}, \ and\ \bibinfo
		{author} {\bibfnamefont {R.}~\bibnamefont {Albert}},\ }\href@noop {}
	{\bibfield  {journal} {\bibinfo  {journal} {Physical Review E}\ }\textbf
		{\bibinfo {volume} {86}},\ \bibinfo {pages} {021924} (\bibinfo {year}
		{2012})}\BibitemShut {NoStop}%
\end{thebibliography}

%

\end{document}